\documentclass[twocolumn,amsmath,amsfonts,amssymb]{revtex4} %,nofootinbib
\usepackage{graphicx}
\def\beq{\begin{equation}}
\def\eeq{\end{equation}}
\def\bea{\begin{eqnarray}}
\def\eea{\end{eqnarray}}

\begin{document}

\title{Dimension of quantum channel of radiation in pure Lovelock black holes %\thanksref{t1}
}
\author{Behrouz Mirza}
\email{ b.mirza@cc.iut.ac.ir}
\affiliation{Department of Physics, Isfahan University of Technology, Isfahan, 84156-83111, Iran}
\author{Fatemeh Oboudiat}
\email{f.oboudiat@ph.iut.ac.ir}
\affiliation{Department of Physics, Isfahan University of Technology, Isfahan, 84156-83111, Iran}
\author{Somayeh Zare}
\email{ smza721@ph.iut.ac.ir}
\affiliation{Department of Physics, Isfahan University of Technology, Isfahan, 84156-83111, Iran}


\begin{abstract}
It is known that the emission rate of entropy from a Schwarzschild black hole is exactly the same as that of  a one dimensional quantum channel \cite{Beken}. We calculate the dimension of entropy emission from a $D$ dimensional pure Lovelock black holes. Our results indicate that the dimension of transmission for odd $D$ dimensional space-times is equal to $D$ and for even $D$ dimensional space-times, the dimension of quantum channel becomes  $1+\epsilon (\Lambda)$, where $ \Lambda $ is cosmological constant. It is interesting that  cosmological constant may put some constraint on dimension of quantum channel in even dimensional space-times. The effect of Generalized Uncertainty Principle (GUP) on the dimension of transmission of entropy for a Schwarzschild black hole is also investigated.
\end{abstract}
\maketitle

%%%%%%%%%%%%%%%%%%%%%%%%%%%%%%%%%%%%%%%%%%%%%%%%%%%%%
\section{Introduction}
Hawking published his idea about the radiation from a black hole as a tunneling process due to the vacuum fluctuations near the horizon \cite{Haw}. There are several methods to derive the Hawking radiation. In his original derivation, Hawking calculated the Bogoliubove coefficients
between  in and out states in a black hole background.
The Damour-Ruffini method for deriving Hawking radiation is based on the relativistic quantum mechanics in the curved space time \cite{Dam}. Based on Hawking's  first description, Parikh and Wilczek supposed that a virtual particle pair is spontaneously produced near the horizon while the negative energy particle tunnels inward and the positive energy particle escapes to infinity \cite{Parikh}.
Christensen and Fulling also developed a method  based on the calculation of Hawking fluxes \cite{Chris}. They determined the strength of the Hawking radiation flux using the trace of energy-momentum tensor. Also, it is known that the fluxes of  Hawking radiation cancel gravitational anomaly at the horizon \cite{Robinson,Isoa}.
 It is interesting that a black hole  as an entropy emitter behaves as a one-dimensional information channel \cite{Beken}. Recently, Hawking radiation energy and entropy flux of a Schwarzschild black hole are viewed as a one-dimensional quantum channel  using the Landauer transport model \cite{Nation}. This model was first applied to the electrical transport in mesoscopic physics and then used to study thermal transport. According to this model, the one-dimensional quantum channel connects two thermal reservoirs with different temperatures to cause thermal transportation. One  reservoir is the black hole and the other is the environment outside the black hole. The entropy flow is equal to the expected value of the energy-momentum tensor. This method is applied to different black holes such as BTZ, Reissner-Nordstrom, Kerr, Kerr-Newmann, Kaluza-Klein and $5D$ black rings \cite{Zeng,Liu,Kaluza-Klein}. The Hawking radiation from all of these black holes can be seen as a  $1D$ quantum channel. We believe that these results should have a  bearing on the findings of  recent studies that maintain that spacetime might be  two
 dimensional near the Planck scale (for a review see \cite{carlip}). The objective of the present paper  is to calculate the dimensionality of entropy transmission from pure Lovelock black holes \cite{Dadhich,Dadhich1,Dadhich2,Dadhich3}.

 In higher dimensions, the Einstein-Hilbert Lagrangian can be generalized  to new Lagrangians
which exhibit several unique properties. These Lagrangians involve the sum of products of curvature tensors with
the indices contracted in a specific manner. We know that the curvature tensor involves second
derivatives of the metric tensor, $\partial ^{2}g$, and a term involving the product of
curvature tensors that has a cubic term  in $\partial ^{2}g$. Nevertheless, it is possible to construct Lagrangians
%which are polynomials in the curvature tensor
which lead to equations of motion
that only involve up to second derivatives of the dynamical variables. These theories were introduced by Laczos and Lovelock \cite{Lanczos,Lovelock}.

After introducing pure Lovelock black holes, we will investigate the dimensionality of entropy transmission of these black holes. Our results show that in odd space time dimensions, the radiation from a $D$ dimensional pure Lovelock black hole can be described by the $D$ dimensional quantum channel.  In even dimensions, an interesting phenomenon appears, namely,
we obtain a relation between the dimension of the quantum channel and the cosmological constant $\Lambda $.
% and one can see that the dimension of the quantum channel is an integer number if the cosmological constant $\Lambda $ is equal to certain quantized values.
We also investigate the effect of the Generalized Uncertainty Principle (GUP) on the emission rate of entropy from a Schwarzschild black hole. It is believed that the Hisenberg Uncertainty Principle needs to be revised as it is no longer satisfactory for strong gravity regimes. The concept of Generalized Uncertainty Principle (GUP) was first put forward by Mead \cite{Mead} in 1964, and  some other models were proposed later \cite{GUP1}. Many authors have applied GUP to modify Quantum Mechanics and black hole relations \cite{GUP2}. We will show that the emission rate of entropy for a Schwarzschild black hole as given by  GUP deviates by a small factor from that given by the one dimensional channel.\par
%and that assuming  an integer value for the dimension of transmission leads to some constraint on  GUP.\par
The outline of this paper is as follows: In Sec. II, we review the method of Pendry and Bekenstein for maximum emission rate of entropy for the Schwarzschild black hole. Entropy emission rate from a BTZ black hole is considered in Sec. III. In Sec. IV, the dimensionality of radiation from a pure Lovelock black hole is obtained. In Sec. V, we consider the effect of GUP on the emission rate of entropy from a  Schwarzschild black hole.
%%%%%%%%%%%%%%%%%%%%%%%%%%%%%%%%%%%%%%%%%%%%%%%%%%%%%

%%%%%%%%%%%%%%%%%%%%%%%%%%%%%%%%%%%%%%%%%%%%%%%%%%%%%
\section{Entropy emission of Schwarzschild black hole in (D)-dimensional space time}
%%%%%%%%%%%%%%%%%%%%%%%%%%%%%%%%%%%%%%%%%%%%%%%%%%%%%
  In this paper, we use the derivation of Pendry for the entropy emission rate $\dot{S}$  \cite{Pendry}.

 The dimensionality of the transmission system can be inferred from the exponent of the radiation  power $P$ in the expression $\dot{S}(P)$.
 To evaluate this expression, we notice that $\dot{S}=\nu\frac{P}{T}$, where $T$ is temperature, $\nu$ is $\frac{d+1}{d}$ in the flat space time ( $d$ is dimension of space) and equals to another constant in the curved space time \cite{nu}. For unidirectional current of modes $P(T)$  has the following form
%\bea
\begin{eqnarray}
P(T)=\frac{\pi T^{2}}{12}
\end{eqnarray} %%\eea
By eliminating $T$ between $P(T)$ and $\dot{S}(T)$, we have:
\begin{eqnarray} %%\bea
\dot{S}=(\frac{\pi P}{3 })^{1/2}\label{S1}
%\eea
\end{eqnarray}

\noindent By repeating the above analysis in the 3-d space and using the Stefan-Boltzmann law ($P=\frac{\pi ^{2}T^{4}A}{120 }$), we have

\begin{eqnarray}  %%\bea
\dot{S}=\frac{4}{3}\frac{P}{T}=\frac{2}{3}(\frac{2{\pi}^{2}A{P}^{3}}{15})^{1/4}\label{S3}
\end{eqnarray} %%\eea
As in the flat spacetime, the dimensionality of the transmission system can be inferred from the exponent of $P$ in the expression  $\dot{S}(P)$ for the d space dimensions
\begin{eqnarray}  %%%\bea
\dot{S}(P)\propto P^{\frac{d}{d+1}}
%\eea
\end{eqnarray}
So, the entropy transmission in a single photon polarization out of a closed hot black body surface, according to  Eq.(\ref {S3}), is 3-dimensional.

For the radiation from a Schwarzschild black hole of mass $M$ in a $D$ dimensional space time, we must consider that we have $A=\frac{2{\pi}^{\frac{D-1}{2}}}{\Gamma (\frac{D-1}{2})}r_{h}^{D-2}$ and that the Hawking temperature is $T_H=\frac{D-3}{4\pi}r_{h}^{-1}$.
The Stefan-Boltzmann law in $D$ dimensional curved space time
\begin{eqnarray}   %%\bea
P=\sigma _{D}AT^{D}\label{5}
%%\eea
\end{eqnarray}

where, $\sigma _{D}=\bar{\Gamma}\frac{2{\pi}^{\frac{D-2}{2}}}{\Gamma (\frac{D}{2})} k_{B}c(\frac{k_{B}}{hc})^{D-1}\Gamma (D)\zeta (D)$ \cite{Stef} and $\bar{\Gamma}$ is the average of the frequency dependent transmission factor ($\Gamma $) over the Planck spectrum \cite{Page2}.
By Eliminating $r_h$ between the equations for $A$ and $T$ and using Eq.(\ref {5}) and Pendry 's maximum entropy rate for power $P$ $(\dot{S}= \nu \frac{P}{T})$, we will have
\begin{eqnarray}  %%\bea
\dot{S}=\beta  P^{1/2}
%%\eea
\end{eqnarray}

where, $\beta=2{\pi}^{\frac{D-1}{2}}(\frac{\bar{\Gamma}}{\Gamma (\frac{D-1}{2})})^{1/2}(\frac{D-3}{4\pi })^{\frac{D}{2}-1}{\sigma _{D}}^{\frac{1}{2}}$.
According to  Eq.(\ref {S3}), this formula shows that the entropy flow or information out of the Schwarzschild black hole in D-dimensional space time is one-dimensional. It is also known that the Hawking radiation (entropy flow) can be represented by the one dimensional Landauer transport model \cite{Nation}. There are other evidences that are consistent with the idea that  black holes act as a one dimensional quantum channel. It is shown that near the horizon and transverse to the t-x plain ($r = 2M + \frac{x^{2}}{8M}$), dimensional quantities and excitations   are redshifted  away and each outgoing partial wave acts as a (1+1)-dimensional black body wave at Hawking temperature \cite{Carlip1}. Also we can obtain entropy production ratio defined as bellow \cite{Zurek}:
\begin{eqnarray}  %%\bea
R=\frac{ds}{ds_{BH}}=T_{H}\frac{\dot{S}}{\dot{E}}
%%\eea
\end{eqnarray}

where $ds$ is the entropy carried by radiation from black hole to the environment and $ds_{BH}$ is the change of entropy of the black hole through radiation. For radiation into vacuum, the entropy production ratio in 1d space is $50\%$ larger than that of 3d space. It is interesting that we may obtain a general upper bound on the thermal coduct  \cite{Nation}.

\section{Entropy emission of BTZ black hole}
The line element of the BTZ black hole  could be written as \cite{Banados}:
\begin{equation}
ds^{2}=-\Delta dt^{2}+\frac{dr^{2}}{\Delta}+r^{2}(d\phi -\frac{J}{2r^2}dt)^2
\end{equation}
where the laps function is
\begin{equation}
\Delta =-M+\frac{r^2}{l^2}+\frac{J^2}{4r^2}
\end{equation}
and $M,J$ are the mass and angular momentum of the BTZ black hole, respectively.
The Hawking temperature and the area of the event horizon, (when $J=0$) are given by:
\begin{equation}
T_{H}=\frac{1}{4\pi}\frac{d\Delta}{dr}|_{r_h},\qquad A=2\pi r_h \label{10}
\end{equation}
By using Eq.(\ref {5}) $(P=\sigma _{3}AT^3)$,  and the equations for $A$ and $T$ in (\ref{10}) we have:
\begin{equation}
P=\sigma _{3} \frac{r_{h}^{4}}{8\pi ^{2}l^6}
\end{equation}
If we find $r_h$ in terms of $P$ and substitute in equation for temperature yields: $T_{H}=\frac{P^{1/4}}{(2\sigma _{3})^{1/4}(\pi l)^{1/2}}$. Finally by considering Pendry 's maximum entropy rate $(\dot{S}= \nu \frac{P}{T})$, we will obtain
\begin{equation}
\dot{S}=\xi P^{3/4}
\end{equation}
where $\xi=\nu (2\sigma _{3})^{1/4}(\pi l)^{1/2}$.  We conclude that the entropy flow or information out of the BTZ black hole in 3-dimensional spacetime is 3-dimensional.

%%%%%%%%%%%%%%%%%%%%%%%%%%%%%%%%%%%%%%%%%%%%%%%%%%%%%
\section{Entropy emission of Lovelock black holes}
%%%%%%%%%%%%%%%%%%%%%%%%%%%%%%%%%%%%%%%%%%%%%%%%%%%%%
  There are two distinct derivations for the gravitational dynamics. One straightforwardly follows from the Bianchi differential identity, which is the only  geometric relation available. By taking the trace (contraction) of the Bianchi identity, the Einstein equation can be deduced as follows
\begin{eqnarray}  %%\bea
G_{ab}=\kappa T_{ab}-\Lambda g_{ab} && T^{a}_{\quad b;a}=0
%%\eea
\end{eqnarray}
where, $T_{ab}$ is the second rank symmetric tensor that represents the energy momentum distribution and $\kappa$ and $\Lambda$ are constants.
In the second derivation of the gravitational dynamics, the variation on the Einstein-Hilbert Lagrangian leads to the divergence free Einstein tensor, $G_{ab}$, and, subsequently, to the Einstein equation.
The generalized Lagrangian in higher dimensions is a Lovelock Lagrangian, which takes the following form
\begin{eqnarray}  %%\bea
L=\sqrt{-g}(\alpha_{0}+\alpha_{1}R+\alpha_{2}(R^2+R_{\alpha \beta \mu \nu }R^{\alpha \beta \mu \nu }\\-4R_{\mu \nu }R^{\mu \nu })+\alpha_{3}O(R^{3}))\nonumber
%%\eea
\end{eqnarray}
where, $\alpha_{0}$ corresponds to the cosmological constant $(\Lambda )$, $\alpha_{1}$ is a coupling constant that represents the standard Einstein-Hilbert term and the second order term i.e. $L_{2}=R^2+R_{\alpha \beta \mu \nu }R^{\alpha \beta \mu \nu }-4R_{\mu \nu }R^{\mu \nu }$ is precisely the quadratic Gauss-Bonnet term. $\alpha_{n}$ with $n\geq 2$ are designates the coupling constants of the higher order terms that represent ultraviolet corrections to Einstein theory.

The theory involving only the first three terms is known as the Einstein-Gauss-Bonnet (EGB) theory. The remarkable property of the Lovelock Lagrangian is that it dose not contain the squares of the second derivative so, the equation of motion remains quasi-linear. An analogue of the Riemann tensor has been introduced, which is a polynomial in the Riemann curvature. It also has been shown that it is possible to derive an analogue of $G_{ab}$ for higher dimensional gravities. The analogue of $G_{ab}$, i.e. a divergence free $H_{ab}$, can be obtained by a variation upon the Lovelock Lagrangian and by tracing the Bianchi derivative of a tensor which is homogeneous quadratic in the Riemann curvature \cite{Dadhich}.
The Lovelock curvature polynomial can be defined as follows
\begin{eqnarray}  %%\bea
\nonumber R_{abcd}^{(n)}&=&F_{abcd}^{(n)}-\frac{n-1}{n(D-1)(D-2)}F^{(n)}(g_{ac}g_{bd}-g_{ad}g_{bc}),\\
F_{abcd}^{(n)}&=&Q_{ab}^{\quad mn}R_{cdmn}
%%\eea
\end{eqnarray}
\begin{eqnarray}  %%\bea
\nonumber Q^{ab}_{\quad cd}&=&\delta ^{ab{a}_{1}{b}_{1}...{a}_{n}{b}_{n}}_{cd {c}_{1}{d}_{1}...{c}_{n}{d}_{n}}R_{{a}_{1}{b}_{1}}^{\quad {c}_{1}{d}_{1}}...R_{{a}_{n}{b}_{n}}^{\quad {c}_{n}{d}_{n}},\\
Q_{\quad ;d}^{abcd}&=&0
%%\eea
\end{eqnarray}
The analogue of $n^{th}$ order Einstein tensor is as follows
\begin{eqnarray}  %%\bea
G^{(n)}_{ab}=n(R^{(n)}_{ab}-\frac{1}{2}R^{(n)}g_{ab})
%%\eea
\end{eqnarray}
and
\begin{eqnarray} %%\bea
R^{(n)}=\frac{D-2n}{n(D-2)}F^{(n)}
%%\eea
\end{eqnarray}
where, $F^{(n)}$ is the Lovelock action polynomial. The dynamics of gravity in higher dimensions (the Einstein-Lovelock equation) would be
\begin{eqnarray}  %%\bea
\sum\alpha_{n}G^{(n)}_{ab}=0
%%\eea
\end{eqnarray}
where, $\alpha_0=\Lambda$, $G^{(0)}_{ab}=g_{ab}$, $G^{(1)}_{ab}=G_{ab}$ is the Einstein tensor and $G^{(2)}_{ab}=H_{ab}$ is the Gauss-Bonnet analogue. We shall focus on the pure Lovelock equation given by
\begin{eqnarray}  %%\bea
G^{(n)}_{ab}=\Lambda g_{ab}
%%\eea
\end{eqnarray}
The spherically vacuum solution of the above equation is
\begin{eqnarray}  %%\bea
ds^{2}=f(r)dt^{2}-\frac{1}{f(r)}dr^{2}-r^{2}d\Omega^{2}_{D-2}
%%\eea
\end{eqnarray}
with
\begin{eqnarray}  %%\bea
f(r)=1-r^{2}(\Lambda +\frac{\mu }{r^{D-1}})^{1/n}
%%\eea
\end{eqnarray}
where, $\mu$ is the black hole mass parameter, and $D$ is the spacetime dimension \cite{Dadhich,Dadhich2}. It should be noted that $G^{(n)}_{ab}$ is non-zero in dimension $D>2n$, hence the critical odd and even dimension are $D=2n+1$ and $D=2n+2$ respectively. The black hole temperature is computed by evaluating the expression $T=\frac{1}{4\pi }f^{\prime}(r_h)$ to obtain \cite{Dadhich3}
\begin{eqnarray}  %%\bea
   T=\left\{
         \begin{array}{cc}
           \frac{\mu -1}{2\pi r_h}  & D=2n+1 \\
           \frac{1}{2\pi }[-\frac{1}{r_h}+\frac{D-1}{D-2}\frac{\mu }{r_{h}^{2}}] & D=2n+2 \\
         \end{array}\right.\label{temp}
    %%\eea
    \end{eqnarray}
Using $f(r)\mid _{r=r_h}=0$, we can initially obtain the following expressions for $\mu $
\begin{eqnarray}  %%\bea
\mu =\left\{
         \begin{array}{cc}
           1-\Lambda r_{h}^{D-1}  & D=2n+1 \\
           r_{h}-\Lambda r_{h}^{D-1} & D=2n+2 \\
         \end{array}\right.
%%\eea
\end{eqnarray}
If we substitute the above expressions for $\mu$ in Eq.(\ref {temp}), we will have
\begin{eqnarray} %%\bea
   T=\left\{
         \begin{array}{cc}
           \frac{-\Lambda}{2\pi}r_{h}^{D-2}  & D=2n+1 \\
           \frac{1}{2\pi }[\frac{1-\Lambda(D-1)r_{h}^{D-2}}{(D-2)r_h}] & D=2n+2\\
         \end{array}\right.\label{temp2}
 %%\eea
 \end{eqnarray}
First it should be noted that in odd dimensions the cosmological constant must be negative $\Lambda =-\frac{(D-1)(D-2)}{2l^2}$. We will calculate the dimensionality of the transmission of entropy or information, for the pure Lovelock black hole. For $D=odd$, substituting the Hawking temperature of the pure Lovelock black hole, i.e, Eq.(\ref {temp2}), and the area of the black hole horizon
$A=\frac{2{\pi}^{\frac{D-1}{2}}}{\Gamma (\frac{D-1}{2})}r_{h}^{D-2}$, in the D-dimensional version of the Stefan-Boltzman law
$(P=\sigma _{D}AT^D)$ yields the following expression for $r_h$ in terms of $P$
\begin{eqnarray}  %%\bea
r_h=\alpha P^{1/{(D-2)(D+1)}}
%%\eea
\end{eqnarray}
where, $\alpha=\left(\frac{1}{\sigma _{D}}(\frac{4\pi l^2}{(D-1)(D-2)})^{D}\frac{\Gamma (\frac{D-1}{2})}{2{\pi}^{\frac{D-1}{2}}}\right)^{1/{(D-2)(D+1)}}$.\par
Therefore, the Hawking temperature of the pure Lovelock black hole for $D=odd$  can be written as follows
\begin{eqnarray}  %%\bea
T=\frac{(D-1)(D-2)}{4\pi l^2}\alpha ^{D-2}P^{1/{D+1}}
%%\eea
\end{eqnarray}
 Finally, the Pendry's maximum entropy rate $(\dot{S}(P))$ for the pure Lovelock black hole can be obtained as
\begin{eqnarray}  %%\bea
\dot{S}=\nu  \frac{P}{T}=\nu\frac{4\pi l^2}{(D-1)(D-2)}\alpha ^{2-D}P^{\frac{D}{D+1}}\label{odd}
%%\eea
\end{eqnarray}
Eq.(\ref {odd}) shows that in an odd $D$ dimensional spacetime, entropy emission from a pure Lovelock black hole is $D$ dimensional.
In odd dimension pure Lovelock gravity has a similar behavior as a BTZ black hole \cite{Dadhich1}. In both of BTZ black hole and pure lovelock black hole in odd dimensions the dimension of quantum channel is obtained to be the same as spacetime dimensions.
\\For $D=even$, we will have the following equation for the radiation power $P$
\begin{eqnarray} %%\bea
P=\gamma r_{h}^{-2}(1-\Lambda(D-1)r_{h}^{D-2})^{D}
%%\eea
\end{eqnarray}
where, $\gamma =\frac{\bar{\Gamma}\sigma _{D}}{(2\pi)^{D}{(D-2)}^{D}}\frac{2{\pi}^{\frac{D-1}{2}}}{\Gamma (\frac{D-1}{2})}$.
From the above equation, we can obtain the following relation for $r_h$ in terms of $P$
\begin{eqnarray}  %%\bea
r_{h}=({\gamma }^{-1}P)^{-1/(2+\epsilon )}
%%\eea
\end{eqnarray}
where, $\epsilon =-D \ln_{r_{h}}(1-\Lambda(D-1)r_{h}^{D-2})$.
Therefore, we can write $(\dot{S}(P))$ as follows
\begin{eqnarray} %%\bea
\dot{S}=\nu  \frac{P}{T}=\nu (2\pi )(D-2)\gamma ^{\frac{D+\epsilon }{D(\epsilon +2)}}P^{\frac{1+\epsilon }{(1+\epsilon)+1}}\label{SSSS}
%%\eea
\end{eqnarray}

 Since in even dimensions, the cosmological constant can be negative, as well as radiation from the event horizon, the radiation from the cosmological horizon should be considered. In this way, we may calculate the dimension of quantum channel related to the cosmological horizon that leads to $1+\epsilon =1-D \ln_{r_{c}}(1-\Lambda(D-1)r_{c}^{D-2})$, where $r_c$ is the position of cosmological horizon. According to the above equation, entropy emission from a pure Lovelock black hole in even dimensions is different from that of a Schwarzschild black hole by a factor of $\epsilon$. Since Schwarzschild black hole in four dimensional spacetime is pure Lovelock case with $n=1$ and $\Lambda =0$ \cite{Dadhich1} the dimension of quantum channel for Schwarzschild black hole is $1+\varepsilon$ with $\varepsilon(\Lambda =0)=0$. So equation (\ref{SSSS}) is consistent with the result of section II about Schwarzschild black hole. Also the channel should be exactly one dimensional for all even dimensional Lovelock black holes with $\Lambda =0$. It is interesting that the value of the cosmological constant may put some constraints on the dimension of the quantum channel  or vies versa: $\Lambda_{D}=\frac{1-r_{h}^{\frac{-\epsilon}{D}}}{(D-1)r_{h}^{D-2}}$.  By considering  a particular value for the cosmological constant, one should be able to have black holes of various sizes and specific values of the dimension of radiation quantum channel.  As a special case, If we assume that the dimensionality of transmission is an integer number; $1+\epsilon =a=2,4,...,D$,  then the connection between the cosmological constant and the dimension of radiation quantum channel can be obtained as follows: $\Lambda_{D}=\frac{1-r_{h}^{\frac{1-a}{D}}}{(D-1)r_{h}^{D-2}}$.
 %Assuming  that the dimensionality of transmission is an integer number; $1+\epsilon =a=2,4,...,D$ leads to a quantized value for $\Lambda $ which is given by $\Lambda_{D}=\frac{1-r_{h}^{\frac{1-a}{D}}}{(D-1)r_{h}^{D-2}}$.
 This analysis suggest a possible connection  between the cosmological constant and a quantum theory of gravity. Recent numerical and analytical studies  suggest that the  dimension of  spacetime may be spontaneously reduced to two dimensions near the Planck scale \cite{carlip}. For even dimensional spacetimes, our  results support these suggestions. However, odd dimensional pure Lovelock black holes are new objects with a possibly different behavior near the planck scale. It is interesting to  further study odd dimensional pure Lovelock black holes as they may really stay $D$ dimensional near the planck scale.

%%%%%%%%%%%%%%%%%%%%%%%%%%%%%%%%%%%%%%%%%%%%%%%%%%%%%
\section{Modification to black hole radiation due to GUP}
%%%%%%%%%%%%%%%%%%%%%%%%%%%%%%%%%%%%%%%%%%%%%%%%%%%%%
 One off the predictions of quantum gravity theories such as string theory, loop gravity, doubly special relativity etc, is the existence of a minimum measurable length or area. This has led to the so called Generalized Uncertainty Principle or GUP \cite{Mead,GUP1,GUP2}. To study the modification to the black hole radiation due to GUP, we should notice two corrections; the modification to Stefan-Boltzmann radiation law and the  black hole temperature correction. Using the modified commutation relation of the form
\begin{eqnarray}  %%\bea
[x,p]=i(1+bp^{2})
%%\eea
\end{eqnarray}
we have the modified Stefan-Boltzmann law for the single photon radiation channel as follows \cite{Stefan-Boltzmann}
\begin{eqnarray} %%\bea
\frac{P}{A}=\alpha T^{4}-b\beta T^{6}
%%\eea
\end{eqnarray}
where
\begin{eqnarray} %%\bea
\nonumber\alpha &=& \frac{\overline{\Gamma}\pi^{2}}{120}  \\     \beta &=& \frac{\overline{\Gamma}}{24}(\frac{18.5\pi^{2}}{15}+\frac{34\pi^{4}}{63})
%%\eea
\end{eqnarray}
\\For temperature correction along the lines of \cite{temp}, we have
 \begin{eqnarray} %%\bea
T_{H}=\frac{1}{8\pi M}(1+\frac{b}{16M^{2}})
%%\eea
\end{eqnarray}
where, M is the black hole mass. By substituting $A=4\pi r_{s}^{2}=4\pi(2M)^{2}$, we can eliminate $M$ between radiation power $P$ and entropy current $\dot{S}$ (in the form of $\dot{S}=\nu \frac{P}{T}$) and calculate $\dot{S}(P)$ as follows
\begin{eqnarray} %%\bea
\dot{S}&=&(\frac{\nu^{2}\overline{\Gamma}\pi P}{480})^{1/2}[1+\frac{b}{\overline{\Gamma}}(4.97\pi-\frac{5.78}{\pi})P]\\
&=&(\frac{\nu^{2}\overline{\Gamma}\pi }{480})^{1/2}P^{\frac{1+\epsilon}{(1+\epsilon)+1}}
%%\eea
\end{eqnarray}
where
\begin{eqnarray} %%\bea
\epsilon=-\frac{P}{\ln P}\frac{b}{\overline{\Gamma}}(4.97\pi-\frac{5.78}{\pi})
%%\eea
\end{eqnarray}
As a special case, if we assume an integer dimension for the transmission radiation, i.e. $1+\epsilon =D=1,2,3,4,$ we can obtain a quantized value for b as follows
\begin{eqnarray} %%\bea
b=\frac{\overline{\Gamma}(D-1)\ln P}{P(\frac{5.78}{\pi}-4.97\pi)}
%%\eea
\end{eqnarray}
So, we have a relation for b in terms of the radiation power P and the dimensionality of entropy transmission $D$.
So, an integer dimension of the entropy emission imposes some constraints on GUP.
%The physical significance of the dimension of the quantum channel is unknown.
\ \
%%%%%%%%%%%%%%%%%%%%%%%%%%%%%%%%%%%%%%%%%%%%%%%%%%%%%
\section{Concluding remarks}
%%%%%%%%%%%%%%%%%%%%%%%%%%%%%%%%%%%%%%%%%%%%%%%%%%%%%
Entropy emission from a $D$-dimensional Schwarzschild black hole behaves as a one dimensional quantum channel. In this paper, we considered the dimension of entropy emission from $D$ dimensional pure Lovelock black holes. It was seen that in odd space time dimensions, the quantum channel of radiation is $D$ dimensional. In even dimensions, %assuming an integer
cosmological constant put some constraints on the dimension of quantum channel. However, pure Lovelock black holes are the first example that
radiate through D-dimensional quantum channel. Although Lovelock black holes are
not in the framework of General Relativity but their study from this point of view has
some new results.
% such as quantisation of the cosmological constant.
 Finally, we  considered the effect of the  Generalized Uncertainty Principle (GUP) on the emission rate of entropy for a Schwarzschild black hole to find that it may deviate from its one dimensional behavior. It should be noted that we do not know if the dimension of quantum channel should be an integer or might be other values. Assuming an integer number for the dimension of quantum channel leads to some constraints on cosmological constant.

It may be concluded that a  relation might exist between the  dimension of quantum channel and recent findings about the dimensional reduction of spacetime near the Planck scale \cite{carlip}. We may expect that for pure Lovelock black holes such dimensional reductions does not appear in odd spacetime dimensions.

%%%%%%%%%%%%%%%%%%%%%%%%%%%%%%%%%%%%%%%%%%%%%%%%%%%%%

%%%%%%%%%%%%%%%%%%%%%%%%%%%%%%%%%%%%%%%%%%%%%%%%%%%%%%%%%%%%%%%%%%%%%%%%%%%%

\end{document}